\documentclass[prd,nofootinbib,12pt]{revtex4}

\usepackage{lineno,hyperref}
\usepackage{amsfonts,amssymb,amsmath}
\usepackage{epsfig}
\usepackage{mathrsfs}
\usepackage{amssymb}
\usepackage{graphicx}
\usepackage{amsmath}
\usepackage{graphicx}

\begin{document}

\title{Gravitational microlensing in Verlinde's emergent gravity}

\author{Lei-Hua Liu$^{1}$}
\email{L.Liu1@uu.nl}
\author{Tomislav Prokopec$^{1}$}
\email{t.prokopec@uu.nl}
\affiliation{1. Institute for Theoretical Physics, Spinoza Institute and
the Center for Extreme Matter and Emergent Phenomena (EMME$\Phi$), Utrecht University,
Buys Ballot Building,
Princetonplein 5, 3584 CC Utrecht, the Netherlands}

\begin{abstract}

We propose gravitational microlensing as a way of testing the emergent gravity theory recently proposed by
Eric Verlinde~\cite{Verlinde:2016toy}. 
We consider two limiting cases: the dark mass of maximally anisotropic pressures (Case I) and of isotropic pressures (Case II).
Our analysis of perihelion advancement of a planet shows that only Case I yields a viable theory. 
In this case the metric outside a star of mass $M_*$ can be modeled by that of a point-like global monopole whose mass
is $M_*$ and a deficit angle $\Delta = \sqrt{(2GH_0M_*)/(3c^3)}$, where $H_0$ is the Hubble rate and $G$ the Newton constant.
This deficit angle can be used to test the theory since light exhibits additional bending around stars given by,
$\alpha_D\approx -\pi\Delta/2$. This angle is independent on the distance from the star and it affects equally light and massive particles.
The effect is too small to be measurable today, but should be within reach of the next generation of high resolution telescopes.
Finally we note that the advancement of periastron of a planet orbiting around a star or black hole, which equals $\pi\Delta$
 per period,  can be also used to test the theory.
\end{abstract}

\maketitle


\section{Global monopole metric}
\label{Global monopole metric}

 In a recent paper Eric Verlinde~\cite{Verlinde:2016toy} has proposed a novel emergent gravity theory.
The most important claim of the theory is that dark matter has no particle origin but instead it is
an emergent manifestation in modified gravity. Assuming spherical symmetry Verlinde obtains,
\begin{equation}
 \int_0^r\frac{G M_D(r^\prime)^2}{{r^\prime}^2}dr^\prime = \frac{cH_0M_B(r)r}{6}
\,,
\label{Verlinde's formula}
\end{equation}
where $H_0=2.36\times 10^{-18}~{\rm s^{-1}}\simeq \sqrt{\Lambda/3}$
is the current Hubble parameter, $\Lambda$ is the cosmological
constant (whose value is determined by the current dark energy density),
$G=6.674\times 10^{-11}~{\rm m^3/(kg s^2)}$ is the Newton's constant,
$c\approx 3\times 10^8~{\rm m/s}$ is the speed of light,  $M_B(r)$ ($M_D(r)$)
is the baryonic mass (dark mass) inside a sphere of
radius $r$.

Eq.~(\ref{Verlinde's formula}) implies that for a star of uniform density
$\rho_*$, $M_*=\frac{4\pi}{3}r^3 \rho_*$, inside the star,
\begin{equation}
  M_D(r) = \sqrt{\frac{2cH_0M_*r^5}{3GR_*^3}}\propto r^{5/2}
\,,\qquad r<R_*
\,,
\label{MD inside star}
\end{equation}
where $R_*$ denotes star's radius. On the other hand, outside the star, $M_D\propto r$, and we have,
\begin{equation}
  M_D(r) = \sqrt{\frac{cH_0M_*}{6G}}\!\times \!r
\,,\qquad r\geq R_*
\,.
\label{MD outside star}
\end{equation}

 The main goal of this paper is to construct the metric tensor that consistently incorporates~(\ref{Verlinde's formula})
within the Verlinde's emergent gravity theory and to investigate how that metric can be used to
test the theory. The fundamental assumption we make is that the theory admits metric formulation
that can be obtained by solving suitably modified Einstein's equations~(\ref{Einstein equation}).
 In the Appendix we perform a detailed analysis of such a theory. Unfortunately, we do not have all of the information
needed to fully specify the metric. A reasonable assumption is that the modified stress energy tensor is diagonal,
$T_\mu^{\;\nu}={\rm diag}[-\rho,P_r,P_\theta,P_\varphi]$, see~(\ref{Tmn}).
In the weak gravitational field regime (which is of our principal concern here) that should be justified. This leaves us with
four unknown functions: energy density $\rho$ (which we can determine from~(\ref{Verlinde's formula}))
and three unknown pressures: $P_r,P_\theta,P_\varphi$. For spherically symmetric  mass distribution the two angular
pressures must be equal, $P_\theta=P_\varphi\equiv P_\perp$. The remaining pressures are unknown, but are nevertheless
tightly constrained by the TOV equation~(\ref{TOV equation}), however not enough to be completely specifiable.
Rather than attempting to extend Verlinde's theory to obtain a relationship between the energy density and pressures, here we consider two simple and plausible {\it Ans\"atze}:
\begin{enumerate}
\item[] {\bf Case I:} Field-like dark mass: $P_\perp = 0$;
\item[] {\bf Case II:} Particle-like dark mass: $P_\perp = P_r\equiv P$.
\end{enumerate}
In addition, in Case II, we assume that inside a star (where except at very small radii baryonic contribution dominates)
baryonic matter is non-relativistic, and hence $P_B\ll \rho_B$, implying also $P\ll \rho$.

The extensive analysis in the Appendix
({\it cf.} Eqs.~(\ref{Gtt:4}), (\ref{Case I: alpha}) and (\ref{Case II: alpha:2}))
shows  that the metric tensor is of the form,
\begin{equation}
 ds^2 = -\left(\!1\!-\!\Delta\!-\!\frac{1\!-\!w^\prime}{2}\frac{H_0^2r^2}{c^2} \!-\!\frac{2GM_*}{c^2r}\right)
            \left(\frac{r}{r_H}\right)^{(1+w^\prime)\Delta}\!\!c^2dt^2
           + \frac{dr^2}{1\!-\!\Delta\!-\!\frac{H_0^2r^2}{c^2}\! -\!\frac{2GM_*}{c^2r}}
            + r^2d\Omega_2^2
\,.
\label{global monopole metric}
\end{equation}
where $d\Omega_2^2=d\theta^2+\sin^2(\theta)d\varphi^2$ is the metric of the two-dimensional unit sphere
($\theta\in[0,\pi]$, $\varphi\in[0,2\pi)$),
$w^\prime=P/\rho=-1$ is the equation of state parameter for Case I ($P_r=P,P_\perp=0$)  and 
$w^\prime =0$ for Case II ($P_r=P=P_\perp$).
This then implies that outside the star for Case I the metric can be written
as that of a point-like global monopole on de Sitter background,~\footnote{Global monopoles are topological
solutions of classical equations of motion of a scalar field theory with $3$ real scalar fields,
 $\vec\Phi=(\Phi_1,\Phi_2,\Phi_3)^T$
whose Lagrangian is $O(3)$ symmetric and whose potential exhibits
a spontaneous symmetry breaking, $V(\vec \Phi)=(\lambda/4)(\vec\Phi^T\cdot \vec\Phi -\Phi_0^2)^2$. One can show
that in this case the solution with topological charge 1~\cite{Marunovic:2014hla} will backreact on the metric
such to induce a solid deficit angle, $\Delta=8\pi G\Phi_0^2$ ($c=1$), see e.g.~\cite{Barriola:1989hx,Marunovic:2015wha}.
From the gravitational point of view compact star-like dense objects (black hole mimickers)
built out of topologically charged scalar matter~\cite{Marunovic:2013eka} resemble ordinary stars in Verlinde's theory.}
and $\Delta$ is the deficit solid angle defined by,
\begin{equation}
        \Delta = \sqrt{\frac{2GH_0M_*}{3c^3}}
\,.
\label{deficit angle}
\end{equation}
That $\Delta$ in~(\ref{deficit angle}) indeed represents a deficit solid angle that cannot be removed by
a coordinate transformation can be shown as follows. Observe firstly that the volume (surface area) of
a two sphere of radius $r$ is $\Omega(S^2(r))=4\pi r^2$, which defines the coordinate $r$ (these coordinates
 are similar to those used in the Schwarzschild metric). Now, one can try to remove $\Delta$ by the following coordinate
transformations,
\begin{equation}
   \tilde r =\frac{r}{\sqrt{1-\Delta}}
\,,\qquad    \tilde t =(1-\Delta)^{\frac12[1+(1+w^\prime)\Delta]}\times t
\,,\qquad
 (1-\Delta)d\Omega_2^2 = d\tilde\Omega_2^2
\label{coordinate transformation}
\end{equation}
after which $\Delta$ seems to disappear from the metric~(\ref{global monopole metric}). Indeed, the equivalent metric is,
\begin{equation}
  ds^2 = -\left(1-\frac{1-w^\prime}{2}\frac{H_0^2\tilde r^2}{c^2} -\frac{2G\tilde M_*}{c^2\tilde r}\right)
                         \Big(\frac{\tilde r}{r_H}\Big)^{(1+w^\prime)\Delta}c^2d\tilde t^{\,2}
           + \frac{d\tilde r^2}{1-\frac{H_0^2r^2}{c^2} -\frac{2G\tilde M_*}{c^2r}}
            + \tilde r^2d\tilde \Omega_2^2
 \,,
\label{global monopole metric:2}
\end{equation}
where
\begin{equation}
 \tilde M_* = \frac{M_*}{(1-\Delta)^{3/2}}
 \,.
\label{global monopole metric:3}
\end{equation}
However, $\Delta$ does not entirely disappear since in the new coordinates,
\begin{equation}
d\tilde \Omega_2^2 = d\tilde \theta^2 + \sin^2\bigg(\frac{\tilde\theta}{\sqrt{1-\Delta}}\bigg)d\tilde\varphi^2
\label{angular metric tilde}
\end{equation}
and $\tilde \varphi$ and $\tilde\theta$ take values in the intervals,
\begin{equation}
  \tilde \theta\in[0,\pi\sqrt{1-\Delta}\,]\,,\qquad
 \tilde \varphi\in[0,2\pi\sqrt{1-\Delta}\,)
\,.
\label{new angular coordinates}
\end{equation}
It is an easy exercise to calculate the surface area of the two dimensional sphere of radius $\tilde r$
in these new coordinates,
\begin{equation}
 \Omega(S^2(\tilde r)) = 4\pi (1-\Delta)\tilde r^2
\,.
\end{equation}
From this result it is obvious that the sphere
contains a solid angle deficit of, $\delta\Omega = -4\pi\Delta$, completing the proof.
In the following section we discuss the physical significance of this result.


\section{Gravitational lensing}
\label{Gravitational lensing}

\begin{figure}[h!]
 \centering
  \includegraphics[width=0.6\textwidth]{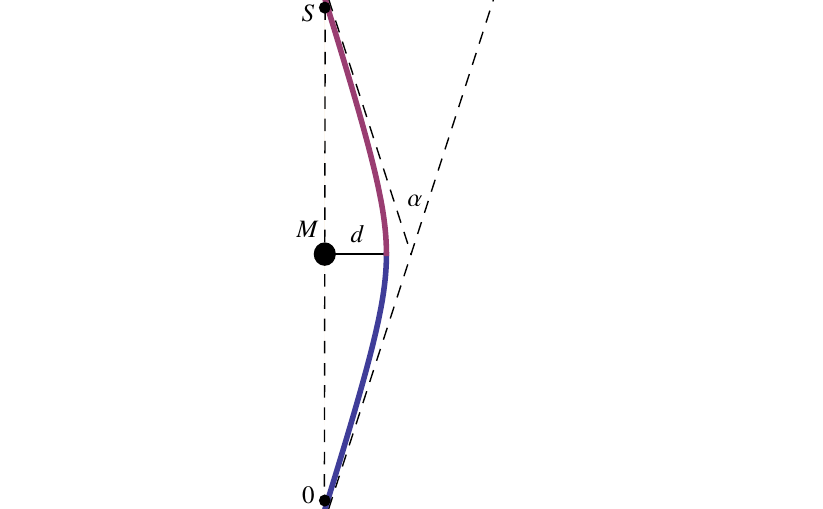}
 \caption{Light deflection around a star of mass $M_*=M$. The closest distance to the center of the star is $d$.
The deflection angle $\alpha$ can be calculated by integrating~(\ref{lensing formula}) along the path of light
from the source $S$ to the observer $O$.}
\label{figure two}
 \end{figure}

 In this section we consider the lensing in a metric given by~(\ref{global monopole metric}) and~(\ref{global monopole metric:2}).
The usual weak (linearised) lensing formula for the deflection angle (in radians),
\begin{equation}
 \alpha = -\frac{1}{c^2}\int \nabla_\perp\big[ \phi(\vec x) +\psi(\vec x)\big]d\ell
\,,
\label{lensing formula}
\end{equation}
where $\ell$ is the path along the light geodesic (from the source to the observer, see figure~\ref{figure two}),
$\nabla_\perp$ is the gradient operator in the plane orthogonal to the propagation of light and
$\phi$ and $\psi$ are the two gravitational potentials (corresponding to the $g_{00}$ and $g_{rr}$ metric perturbations).
Outside the star these potentials can be read off from~(\ref{global monopole metric}),
\begin{equation}
 \phi(r) = -\frac{GM_*}{r} -\frac{1-w^\prime}{2}\frac{H_0^2}{2}r^2
                           +\frac{c^2(1+w^\prime) \Delta}{2}\ln\Big(\frac{r}{r_H}\Big) -\frac{c^2\Delta}{2}
\,,\quad
 \psi(r) = -\frac{GM_*}{r} -\frac{H_0^2}{2}r^2 -\frac{c^2\Delta}{2}
\,.
\label{total potential}
\end{equation}
The lensing formula~(\ref{lensing formula})
can be used for the first three parts of the potential (the one induced by the star mass, by the Universe's expansion and the logarithmic piece),
but it cannot be used for the constant contribution, $\phi_D=-c^2\Delta/2$, from the dark mass simply because, $\nabla_\perp\phi_D=0$
(how to calculate light deflection due to $\phi_D$ is discussed below).
For that reason it is better to use the second form of the metric~(\ref{global monopole metric:2}),
in which case the gravitational potential is,
\begin{equation}
\tilde \phi = -\frac{G\tilde M_*}{\tilde r} -\frac{1-w^\prime}{2}\frac{H_0^2}{2}\tilde r^2
                             +\frac{c^2(1+w^\prime)\Delta}{2}\ln\Big(\frac{\tilde r}{r_H}\Big)
\,,\quad
\tilde \psi = -\frac{G\tilde M_*}{\tilde r} -\frac{H_0^2}{2}\tilde r^2
\label{total potential}
\end{equation}
Inserting this into~(\ref{lensing formula}) gives for the lensing angle,
\begin{equation}
 \alpha_1 = -\frac{4GM_*}{c^2(1-\Delta)^{3/2}d} +\frac{3-w^\prime}{2}\frac{H_0^2\ell d}{c^2}
     - \frac{\pi}{2}(1+w^\prime)\Delta
\,,
\label{lensing formula:2}
\end{equation}
where $\ell=\overline{SO}$ is the distance from the source $S$ to the observer $O$ and $d$ is the closest distance of light from
the star center, see figure~\ref{figure two}.
The first integral was evaluated by assuming that the source and the observer are infinitely far from the star, which
is for small lensing angles an excellent approximation. It is instructive to compare the first two contributions in~(\ref{lensing formula:2}).
Note first that the second part is accurate only when $H_0\ell\ll 1$, because otherwise the linear lensing formula~(\ref{lensing formula})
fails. This means that the second contribution is $\ll 2H_0d/c^2$. The two contributions will be approximately equal when
$2GM_*/d \sim H_0 d \times (H_0\ell)\ll H_0d$, that means when the distance $d$ expressed in units of
the Schwarzschild radius of the star $R_S$ becomes comparable to the distance expressed in units of the Hubble distance $d_H$, {\it i.e.}
when $d\simeq \sqrt{R_S d_H}\sqrt{d_H/\ell\,}$.

\begin{figure}[h!]
 \centering
 \includegraphics[width=0.3\textwidth]{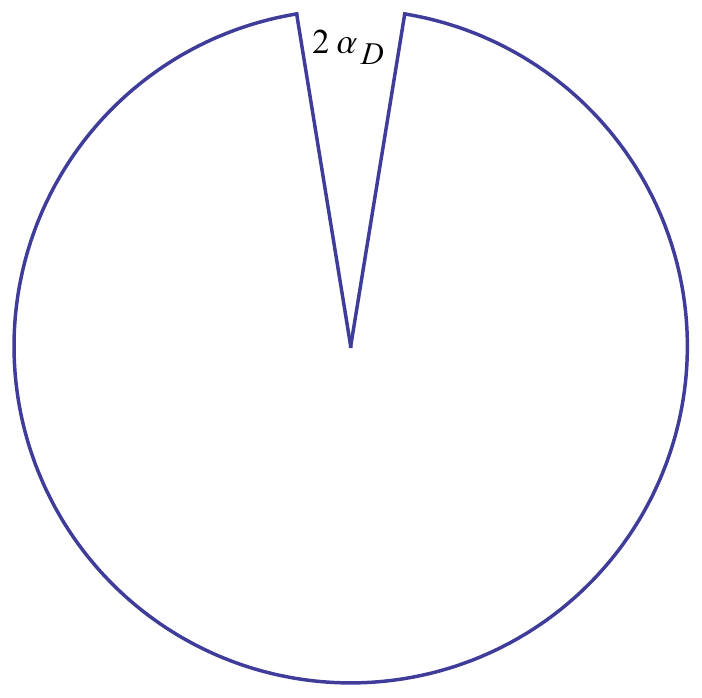}
 \caption{The total deficit angle $2\alpha_D = 2\pi(1-\sqrt{1-\Delta})$ in the equatorial plane $\theta=\pi/2$ around
a star of mass $M_*=M$. The two rays emanating from the origin are identified. Due to the angle deficit,
a light ray from a distant star and observed by a distant observer exhibits a `change' in direction given by $\alpha_D$.}
\label{figure three}
 \end{figure}

 In order to get the lensing generated by the deficit angle $\Delta$, note that it is convenient to assume
that the plane shown in figure~\ref{figure two} corresponds to the equatorial plane $\theta = \pi/2$,
or equivalently $\tilde \theta = (\pi/2)\sqrt{1-\Delta}$. In that plane
the azimuthal angle takes values in the interval, $\tilde \phi\in [0,2\pi\sqrt{1-\Delta})$.
The geometry is flat and can be represented by a plane in which the wedge whose angle
equals to the deficit angle $2\pi[1-\sqrt{1-\Delta}]$ is cut out, and the opposite sides of the wedge are identified,
representing conical geometry (of a global string or a point mass in two spatial directions),
see figure~\ref{figure three}.
Consider now a ray in this conical geometry propagating from a distant source (on one side of the central point
where the star is located)
to a distant observer (on the other side of the central point). Due to the conical geometry,
to a very good approximation that ray will exhibit an angle deflection of one half of the total deficit angle,
{\it i.e.}
\begin{equation}
 \alpha_D = -\pi\big(1-\sqrt{1-\Delta}\,\big)\approx - \frac{\pi}{2}\Delta
\,.
\label{deficit angle dark}
\end{equation}

The total deflection angle is then simply the sum of~(\ref{lensing formula:2}) and~(\ref{deficit angle dark}),
\begin{equation}
 \alpha= \alpha_1+\alpha_D= -\frac{4GM_*}{c^2(1-\Delta)^{3/2}d} +\frac{3-w^\prime}{2}\frac{H_0^2\ell d}{c^2} -\frac{\pi}{2}(1+w^\prime)\Delta
                     -\pi\big(1-\sqrt{1-\Delta}\,\big)
\,.
\label{total deficit angle}
\end{equation}

Let us now see whether $\alpha_D$ can be large enough to be measurable for typical stars.
Consider first our Sun for which,
\begin{equation}
M_*=M_\odot = 1.989\times 10^{30}~{\rm kg}
\,,\quad
R_*=R_\odot =  6.957\times 10^8~{\rm m}
\,,\quad
H_0=2.36\times 10^{-18}~{\rm s^{-1}}\,,
\label{our sun}
\end{equation}
and therefore from~(\ref{total deficit angle}),
\begin{equation}
 \alpha_\odot = \Big(-1.74\frac{R_\odot}{d} +2.26\times {10^{-12}}\frac{3-w^\prime}{4}\frac{d}{R_\odot}\frac{\ell}{d_H}
 - 9 \times {10^{-7}}(2+w^\prime)\Big)[{\rm arcsec}] \qquad {\rm (Sun)}
\,.
\label{total deficit angle:2}
\end{equation}
At first sight the second and third contributions look desperately small.
Note however that, while the first contribution to $\alpha$ in~(\ref{total deficit angle:2}) drops with the distance from the Sun,
the second contribution grows and the third contribution stays constant.
Let us now compare these numbers with the sensitivity of modern observational probes. For example, the ESA's GAIA
mission~\cite{Spagna:2016jnz,GAIA_ESA,Hees:2015ixa}
 (whose purpose is to make 3 dimensional optical image of about one billion stars in the Milky Way by measuring parallaxes of stars)
has a sensitivity of about $20\mu {\rm arcsec}$ for stars of magnitude 15 or larger and $7\mu {\rm arcsec}$ for stars of
magnitude 10 or larger, which is about a factor 20 (7) too low to be able to observe $\alpha_D$ induced by the Sun.
Analogously, the Event Horizon Telescope (EHT)\cite{Doeleman:2009te,EHT}
 -- whose goal is to make a radio map (in wavelengths of about $1~{\rm mm}$)
 of the neighborhood of the Milky Way black hole
located near the galactic center in Sag $A^*$ -- will have angular resolution of about $10~\mu {\rm arcsec}$,
 a factor of 10 above the resolution required to see $\alpha_D$ generated by the Sun. From this analysis we see that
the next generation of even more precise observatories will probably reach the required precision to
be able to see the microlensing induced by the dark potential $\phi_D$ of the Sun.

Let us now have a closer look at our Mikly Way black hole, whose mass is about $M_{\rm BH}=4.3\times 10^6~M_\odot$
and whose Schwarzschild radius is, $R_{\rm S}=2GM_{\rm BH}/c^2= 1\times 10^{10}~{\rm m}$,
\begin{equation}
 \alpha_{\rm BH} = \!- \frac{2R_{\rm S}}{d}[{\rm rad}] + \Big(9.6\times {10^{-18}}\frac{3\!-\!w^\prime}{4}\frac{d}{R_{\rm S}}\frac{\ell}{d_H}
 - 1.7 \times {10^{-3}}(2+w^\prime)\Big)[{\rm arcsec}] \quad ({\rm Sag A^*\; black\; hole})
\,.
\label{total deficit angle:3}
\end{equation}
The result for $\alpha_D$ is within the reach of EHT, but it is not clear whether the EHT mission will be able to measure
so small deflection angles in the vicinity of the black hole.

Note that the angular deflection induced by $\alpha_D$ is equal for all objects, independently on how fast they move.
That means that (massive) objects that move with a speed $v<c$ will exhibit the same `dark' deflection angle $\alpha_D$ as light.
This is not true for the usual gravitational lensing. Indeed, by solving the geodesic equation for ultra-relativistic particles one obtains,
\footnote{
The relevant geodesic equations for the four velocity, $u^\mu=(u^0,\vec u)$, $\vec u = (\vec u_\perp,\vec u_\|)$ are,
\[
\frac{du_\perp}{d\lambda}+\frac{(u^0)^2}{c^2}\nabla_\perp \tilde \phi + \frac{(\vec u)^2}{c^2}\nabla_\perp \tilde \psi = 0
\,,\quad \frac{d\ell}{d\lambda} = \|\vec u\| \equiv u\,,\quad (u^0)^2 = \vec u^2 + c^2 = \gamma^2c^2
\,.
\]
The deflection angle is then, $\alpha\approx u_\perp/u$. By solving the geodesic equations for $u_\perp/u$ 
and adding the deficit angle contribution 
one arrives at~(\ref{deflection massive relativistic particles}).  
}
\begin{eqnarray}
\alpha_m \!&=&\! \Big(1+\frac{1}{\gamma^2}\Big)\Big(-\frac{2GM_*}{c^2(1-\Delta)^{3/2}d} +\frac{1-w^\prime}{2}\frac{H_0^2\ell d}{c^2}
            -\frac{\pi}{2}(1+w^\prime)\Delta
           \Big)
\nonumber\\
   &&-\,\frac{2GM_*}{c^2(1-\Delta)^{3/2}d} +\frac{H_0^2\ell d}{c^2}
 -\pi\big(1-\sqrt{1-\Delta}\,\big)
\label{deflection massive relativistic particles}
\end{eqnarray}
where $\gamma = [1-(v/c)^2]^{-1/2}\gg 1$
and we have neglected the change of velocity parallel to the motion.
In the limit when $\gamma\rightarrow \infty$, Eq.~(\ref{deflection massive relativistic particles}) reduces to 
Eq.~(\ref{total deficit angle}), as it should. 
By comparing the angle of deflection of light~(\ref{total deficit angle})
with that of relativistic particles~(\ref{deflection massive relativistic particles}) (such as cosmic rays which have a large $\gamma$ factor)
one could in principle isolate the component that is independent on the speed of motion, thereby testing the
Verlinde formula~(\ref{Verlinde's formula}). The measurement will not be easy, but it is not impossible.

 Finally, we point out that -- due to the deficit angle -- the dark mass $M_D$ will cause an addition advancement of
perihelion/periastron of a planet, which is per orbit,~\footnote{The result~(\ref{periastron}) follows from the conservation equation
for motion of a planet,
\[
\Big(\frac{dx}{d\varphi}\Big)^2 +(1-\Delta)\frac{c^2L^2}{G^2M_*^2}-2x+(1-\Delta)x^2-\frac{2G^2M_*^2}{c^2L^2}x^3
   = \frac{E^2L^2}{c^2G^2M_*^2}\Big(\frac{x}{x_H}\Big)^{(1+w^\prime)\Delta}
\,,
\]
where $x=L^2/(GM_*r)$. Taking a derivative with respect to $\varphi$ and expanding the solution around the classical one
$x_01+e\cos(\varphi)$, as  $x(\varphi)=x_0+x_1$, where $e$ denotes elipticity, one obtains,
 \[
 x_1  =x_{GR} + \Delta\Big(1+\frac{e}{2}\varphi\sin(\varphi)\Big)  + \frac{E^2L^2}{2c^2G^2M_*^2}
                   \Big(1+\frac{e^2}{2}-\frac{e}{2}\varphi\sin(\varphi)-\frac{e^2}{6}\cos(2\varphi)+{\cal O}(e^3)\Big)
\,,
\]
where $x_{GR}$ denotes the general relativistic correction, and $E$ is the energy per unit mass, which for a circular orbit,
$E^2=c^4\big(1-\frac{G^2M_*^2}{c^2L^2}-\frac{2G^4M_*^4}{c^4L^4}\big)$. 
The advancement of periastron is caused by the term proportional to 
$\varphi\sin(\varphi)$, and the coefficient of that term yields Eq.~(\ref{periastron}).
 }
\begin{equation}
\Delta \phi_{\rm periastron} = \pi\Delta - \pi(1+w^\prime)\Delta\frac{c^2L^2}{2G^2M_*^2}
                   \bigg(1-\frac{G^2M_*^2}{c^2L^2}- \frac{2G^4M_*^4}{c^4L^4}\bigg)   
            \; [{\rm per\; orbit}]
\,,
\label{periastron}
\end{equation}
where $L$ is the angular momentum per unit mass of a planet. 
Note that the method used to derive~(\ref{periastron}) reproduces correctly only the linearised part of the exact expression for the geometric effect,
which equals: $2\pi(1-\sqrt{1-\Delta})$.
For a planet in the solar system the first (geometric) term 
produces a tiny effect.  Indeed,
for any planet in the Solar system the perihelion advancement is, 
$\Delta \phi_{\rm perihelion,1}=\pi \Delta= 1.8\mu{\rm arcsec}$
per orbital period .
This is to be contrasted, for example, with the general relativistic advancement of perihelion of Mercury,
 $\Delta \phi_{\rm GR} = 0.1~{\rm arcsec}$ per period, which is about 50000 times larger.
However, the second term in~(\ref{periastron}) is large. For example, for Mercury
the classical radius is $r_c = L^2/(GM_\odot)=5.5\times 10^{10}~{\rm m}$ and the Schwarzschild radius of the Sun is, 
$r_s=2GM_\odot/c^2\simeq 2950~{\rm m}$ and thus,  
$\Delta \phi_{\rm perihelion,2\; Mercury}\approx -\pi\Delta\times r_c/r_s=-34~{\rm arcsec}$ per orbital period,
which is much larger than the general relativistic effect (for other planets in the Solar system the effect is even larger).
Based on this observation alone, one concludes that either (1) Verlinde's emergent gravity is ruled out, or (2) the dark mass in Verlinde's 
theory is field-like and produces a highly anisotropic pressure (Case II).  If latter is true (which is the one we favour) $w^\prime=-1$ 
in Eq.~(\ref{periastron}), and the only effect that survives is the first (geometric) term which produces a tiny 
effect that is equal for all planets in the solar system. 
Next, recall that the accuracy of current measurements for Mercury is at the level of $\sim 1\%=10^{-2}$,
which is still far above the sensitivity ($\sim 10^{-5}$) required to measure the perihelion advancement generated by
the geometric term in~(\ref{periastron}) (Case~II when $w^\prime=-1$).
For larger stars and large black holes however, the advancement of periastron is much larger and therefore potentially easier
to observe.


\section{Discussion}
\label{Discussion}

 In this letter we construct the metric tensor associated with spherically symmetric distribution of matter in Verlinde's emergent 
gravity~\cite{Verlinde:2016toy}. We consider two cases. In Case I (maximally anisotropic pressures) 
we assume that the angular pressures vanish and (outside the star) obtain the metric tensor of a global monopole. In Case II we assume 
isotropic pressures. The metric tensor containing both cases is given in Eq.~(\ref{global monopole metric}), where different cases are expressed 
through the equation of state parameter, $w^\prime =P/\rho$ (in Case I, $w^\prime =-1$ and in Case II, $w^\prime =0$).
Our analysis of advancement of perihelion/periastron~(\ref{periastron}) shows that only Case I represents a viable model.
In that case the metric~(\ref{global monopole metric}) exhibits a solid angle deficit~(\ref{deficit angle}).
Here we suggest that this deficit angle can be used to test the theory {\it via} gravitational lensing and periastron advancement.
We consider two cases: the metric of the Sun and that of our Galactic black hole.
Even though the effect is tiny the next generation of observatories are expected to reach the angular sensitivity needed
be able to measure it.

Furthermore, we note that the effect of dark mass~(\ref{Verlinde's formula}--\ref{MD outside star})
can be tested by comparing
the deflection angle of relativistic particles~(\ref{deflection massive relativistic particles}) with that of light~(\ref{total deficit angle})
and by precisely measuring the advancement of periastron~(\ref{periastron}) of planets orbiting around stars.

 The extensive experience acquired in microlensing used for tracking down MACHOs and for discovering new (Earth like) extra-Solar planets
might be of use for detecting dark mass~\cite{Wambsganss:2016qpg,Rahvar:2015eba,Mao:2012za}.

Since the effect of dark mass is cumulative, a much larger effect is
generated by galaxies and clusters of galaxies, and a preliminary discussion of that effect (that mimics dark matter)
can be found in the original reference~\cite{Verlinde:2016toy}. 
We point out that after the first version 
of this work several references have appeared~\cite{Brouwer:2016dvq}, \cite{Iorio:2016qzr} that also discuss how to test 
Verlinde's emergent gravity.

\section*{Acknowledgements}

This work is in part supported by the D-ITP consortium, a program of the Netherlands Organization for Scientific
Research (NWO) that is funded by the Dutch Ministry of Education, Culture and Science (OCW). L.L. is funded by a 
Chinese Scholarschip Council (CSC). 

\section*{Appendix}

In this appendix we solve the Einstein equation sourced by the dark mass.
We solve the Einstein equation,
\begin{equation}
 G_{\mu}^{\;\nu} +\delta_{\mu}^{\;\nu}\Lambda = \frac{8\pi G}{c^4} T_{\mu\nu}
\,,
\label{Einstein equation}
\end{equation}
where $\Lambda$ is the (emergent) cosmological constant $G_{\mu}^{\;\nu}$ is the Einstein tensor,
 $\delta_{\mu}^{\;\nu}$ is the Kronecker delta,
and $T_{\mu}^{\;\nu}$ is the energy-momentum tensor which which we assume here to be diagonal and of the form,
\begin{equation}
 T_\mu^{\;\nu} ={\rm diag} \left[-\rho_B(r)-\rho_D(r),P_r,P_\perp,P_\perp\right]
\,,
\label{Tmn}
\end{equation}
where $\rho_B(r)$ is the usual matter contribution, which for a star of constant density $\rho_{*}=(3M_*c^2)/(4\pi R_*^3)$
and radius $R_*$,
\begin{equation}
 \rho_B(r) = \begin{cases} \frac{3M_*c^2}{4\pi R_*^3} & \; {\rm for}\quad r\leq R_* , \cr
                                         0  & \; {\rm for}\quad r>R_* , \cr
                   \end{cases}
\label{rhoB}
\end{equation}
and $\rho_D(r)=[c^2/(4\pi r^2)]dM_{\rm D}/dr$ is the energy density of dark mass~(\ref{MD inside star}--\ref{MD outside star})
given by,
\begin{equation}
 \rho_D(r) = \begin{cases} \frac{5c^2}{4\pi}\sqrt{\frac{cH_0M_*}{6GR_*^3r}} & \; {\rm for}\quad r\leq R_* , \cr
                                          \frac{c^2}{4\pi r^2}\sqrt{\frac{cH_0M_*}{6G}} =\frac{c^4}{8\pi G r^2}\sqrt{\frac{r_S}{3r_H}} &
                                          \; {\rm for}\quad r>R_* , \cr
                   \end{cases}
\label{rhoD}
\end{equation}
where $r_S=2GM/c^2$, $r_H=H_0/c$.
Eq.~(\ref{Tmn}) allows for a contribution from the pressure $P_r=P_r(r)$ and angular pressures,
$P_{\theta}=P_\varphi=P_\perp(r)$ (the two angular pressures must be equal by the symmetry of the problem
and more formally by the angular component of the covariant conservation for $T_{\mu\nu}$).
How to calculate these pressures is not clear from Verlinde's paper~\cite{Verlinde:2016toy}.
Here we shall consider two simple cases, namely:
\begin{itemize}
\item[(I)] Field-like dark mass: $P_r=P(r), P_\perp=0$ and
\item[(II)] Particle-like dark mass: $P_r=P_\perp=P(r)$.
\end{itemize}
Note that in the case of global monopoles~\cite{Barriola:1989hx} and at large distances we have $P_r=-\rho$, and $P_\perp=0$,
which is therefore analogous (not surprisingly) to the field-like case.
Even though we do not know what the pressure in Verlinde's theory is, we
know that the covariant conservation of the energy-momentum tensor must hold,
\begin{equation}
\nabla_\mu T^\mu_{\;\nu}=0
\,.
\label{cov conservation Tmn}
\end{equation}
As we show below this equation contains a very useful information on the problem.

 For a spherically symmetric matter distribution we can assume a static, diagonal metric tensor of the form,
\begin{equation}
 ds^2 = g_{\mu\nu}dx^\mu dx^\nu
   =-{\rm e}^{2\alpha(r)}dt^2+{\rm e}^{2\beta(r)}dr^2+r^2d\Omega_2^2
\,,\quad
           d\Omega_2^2=d\theta^2+\sin^2(\theta)d\varphi^2
\,,
\label{metric ansatz}
\end{equation}
where  $g_{\mu\nu}$ denotes the metric tensor.
Inserting~(\ref{metric ansatz}) into~(\ref{Einstein equation}) gives the following three equations,
\begin{eqnarray}
 G_t^{\;t} \!&=&\! {\rm e}^{-2\beta}\left(\frac{1}{r^2}\!-\!\frac{2\beta^\prime}{r}\right)-\frac{1}{r^2}
                   = -\Lambda -\frac{8\pi G}{c^4}\left(\rho_B(r)+\rho_D(r)\right)
\label{Gtt}
\\
 G_r^{\;r} \!&=&\! {\rm e}^{-2\beta}\left(\frac{1}{r^2}\!+\!\frac{2\alpha^\prime}{r}\right)-\frac{1}{r^2}
                   = -\Lambda +\frac{8\pi G}{c^4}P_r(r)
\label{Grr}
\\
 G_\theta^{\;\theta} \!&=&\! G_\varphi^{\;\varphi}
     ={\rm e}^{-2\beta}\left(\frac{\alpha^\prime\!-\!\beta^\prime}{r}\!+\!\alpha^{\prime\prime}\!+\!(\alpha^\prime)^2
              \!-\!\alpha^\prime\beta^\prime\right)
                   = -\Lambda+\frac{8\pi G}{c^4}P_\perp(r)
\,,
\label{Gthetatheta}
\end{eqnarray}
where $\alpha^\prime=d\alpha/dr, \beta^\prime=d\beta/dr$. The last equation does not provide an independent information
since it is implied by the contracted Bianchi identity, $\nabla_\mu G^\mu_{\;\nu}=0$.

The relevant information in the conservation law~(\ref{cov conservation Tmn})
 is in the $\nu=r$ equation (the $\nu=\theta$ equation tells us that
$P_\theta=P_\varphi=P_\perp$),
\begin{equation}
  P_r^\prime+\Big(\alpha^\prime+\frac{2}{r}\Big)P_r-\frac{2}{r}P_\perp = -\alpha^\prime \rho
 \,.
\label{radial conservation of pressure}
\end{equation}
 This equation and equations~(\ref{Gtt}--\ref{Gthetatheta}) represent the set of equations we ought to solve for
the metric functions $\alpha(r)$ and $\beta(r)$.

The first equation~(\ref{Gtt}) can be solved by noting that
$r^2G_t^{\;t}=\left[r{\rm e}^{-2\beta}\right]^\prime-1$, and hence,
\begin{equation}
 {\rm e}^{-2\beta(r)}=1-\frac{\Lambda}{3}r^2-\frac{8\pi G}{c^4 r}\int_0^r\big(\rho_B(\tilde r)+\rho_D(\tilde r)\big)\tilde r^2 d\tilde r
\,,
\label{Gtt:2}
\end{equation}
where we set an integration constant to zero (this is so because in the absence of matter metric tensor must reduce to
the Minkowski metric). The integral in~(\ref{Gtt:2}) is to be understood such that, at $r=R_*$, $\beta$ is continuous
(as it is required by Eq.~(\ref{Gtt})), {\it i.e. } the integral~(\ref{Gtt:2}) is continuous at $r=r_*$.

 By integrating~(\ref{Gtt:2}) we can obtain the solution for $\beta(r)$ inside the star ($r\leq R_*$),
\begin{equation}
 {\rm e}^{-2\beta(r)} = 1 - \frac{2GM(r)}{c^2r}-\frac{\Lambda}{3}r^2
\,,
\label{beta inside star}
\end{equation}
where
\begin{equation}
M(r) = M_B(r) + M_D(r)
\,,\quad
M_B(r)  = M_*\Big(\frac{r}{R_*}\Big)^3
\,,\quad
M_D(r)  = \sqrt{\frac{2cH_0M_*r^5}{3GR_*^3}}
\quad (r<R_*)
.
\label{beta inside star:2}
\end{equation}

Next let us consider the gravitational field outside the star, $r>R_*$.
We can split the integral in~(\ref{Gtt:2}) to $0\leq r\leq R_*$ and $r>R_*$. The first integral yields a constant,
\begin{equation}
M(R_*)=\frac{4\pi}{c^2}\int_0^{R_*}\big(\rho_B(\tilde r)+\rho_D(\tilde r)\big)\tilde r^2 d\tilde r =
 M_* + \sqrt{\frac{2cH_0M_*}{3G}}R_*  = M_* +\frac{c^2}{2G}\Delta\times R_*
\,.
\label{Gtt:3}
\end{equation}
From this we see that $\rho_D$ inside a star generates a potential that at the surface of the star generates a solid deficit angle,
\begin{equation}
\Delta  \equiv \sqrt{\frac{2H_0GM_*}{3c^3}}
\,.
\label{appendix: deficit angle}
\end{equation}
The required continuity of $\beta$ at $r=R_*$ implies that the deficit angle is inherited by an exterior metric, {\it i.e.}
the interior of a star in Verlinde's emergent gravity has the same effect as the core of a global magnetic monopole and generates
a boundary condition (at star's surface) that corresponds to that of a global monopole.
With this in mind the total integral gives,
\begin{equation}
M(r)=\frac{4\pi}{c^2}\int_{0}^r \big(\rho_B(\tilde r)+\rho_D(\tilde r)\big)\tilde r^2 d\tilde r
           =  M_* +\frac{c^2}{2G}\Delta\times r \,,\qquad (r>R_*)
\,.
\label{Gtt:3}
\end{equation}
When this is inserted into~(\ref{Gtt:2}) one obtains,
\begin{equation}
 {\rm e}^{-2\beta(r)}=1-\Delta-\frac{\Lambda}{3}r^2 - \frac{2GM_*}{c^2r}
\,,\qquad \Delta = \sqrt{\frac{2H_0GM_*}{3c^3}}=\sqrt{\frac{r_S}{3r_H}}
\,,
\label{Gtt:4}
\end{equation}
where $r_S=2GM/c^2$ and $r_H=H_0/c=\sqrt{\Lambda/3}$ denote the Schwarzschild and Hubble radius, respectively.
The main result up to now is that the radial part of the metric tensor in Verlinde's emergent gravity is
generally of the form  form~(\ref{beta inside star}), where inside the star $M(r)$ is given by~(\ref{beta inside star:2}),
while outside the star $M(r)$ is given in~(\ref{Gtt:3}), such that the exterior metric~(\ref{Gtt:4})
exhibits a solid deficit angle $\Delta$ defined in~(\ref{appendix: deficit angle}).
From the main text we know that $\Delta$  signifies a solid deficit angle.
Eqs.~(\ref{beta inside star}) and~(\ref{Gtt:4}) are exact solutions for  $\beta(r)$
when a star is of constant energy density.

\medskip

Next we consider $\alpha=\alpha(r)$ which is determined by Eqs.~(\ref{Grr}) and~(\ref{beta inside star}),
\begin{equation}
\alpha^\prime(r)
        = \frac12\frac{-\frac23\Lambda r+\frac{2GM(r)}{c^2r^2}+\frac{8\pi G}{c^4}r P_r(r)}{1-\frac{\Lambda}{3}r^2-\frac{2GM(r)}{c^2r}}
\,,
\label{Grr:2}
\end{equation}
with $M(r)$ given by Eqs.~(\ref{beta inside star})--\ref{beta inside star:2}) inside the star
 and by Eqs.~(\ref{Gtt:3}--\ref{Gtt:4}) outside the star.
Since Eq.~(\ref{Grr:2}) is sourced by pressure,
one has to consider Eq.~(\ref{Grr:2}) together with the conservation equation~(\ref{radial conservation of pressure}).
It is convenient to use Eq.~(\ref{Grr:2}) to get rid of $\alpha^\prime$ in~(\ref{radial conservation of pressure}), resulting in the
Tolman-Oppenheimer-Volkoff (TOV) equation for hydrostatic equilibrium,
\begin{equation}
  P_r^\prime=-\frac{P_r+\rho}2\!\times\!\frac{-\frac23\Lambda r+ \frac{2GM(r)}{c^2r^2} +\frac{8\pi G}{c^4}r P_r(r)}
                                            {1 - \frac{2GM(r)}{c^2r}-\frac{\Lambda}{3}r^2}
                                  -\frac{2}{r}(P_r-P_\perp)
 \,,
\label{TOV equation}
\end{equation}
which can be solved for $P_r$ (if one knows $P_\perp$).
When the solution of this equation is inserted into~(\ref{Grr:2}) one can solve for $\alpha(r)$.

Let us consider more closely the TOV equation~(\ref{TOV equation}). It is convenient to introduce
dimensionless pressures, $p_r = [8\pi G/c^4] r_H^2 P_r, p_\perp = [8\pi G/c^4] r_H^2 P_\perp$,
energy density  $\tilde\rho =  [8\pi G/c^4] r_H^2 \rho$ and distance, $x=r/r_H$ ($r_H=H_0/c$).
The  TOV equation can then be written as a Riccati differential equation,
\begin{equation}
  \frac{dp}{dx}+\bigg[\frac{2\epsilon}x+\frac12\frac{-2x+\frac{\Delta}{x}+\frac{3\Delta^2}{x^2}+x\tilde\rho(x)}{A(x)}\bigg]p+\frac{x}{2A(x)}p^2
= -\frac{-2x+\frac{\Delta}{x}+\frac{3\Delta^2}{x^2}}{2A(x)}\tilde\rho(x)
 \,,
\label{TOV equation:2}
\end{equation}
where
\begin{equation}
 A(x) = 1-\Delta -x^2 - \frac{3\Delta^2}{x}
\qquad (x>x_*=r/R_*)
\nonumber
\end{equation}
and where
\begin{equation}
 \epsilon = \begin{cases}
                  1 & {\rm for\;\; Case\; I:} \; p_r=p,\; p_\perp =0\,, \cr
                   0 & {\rm for\;\; Case\; II:} \; p_r=p_\perp = p\,. \cr
\end{cases}
\label{epsilon 2 cases}
\end{equation}
We are primarily interested in solving for $\alpha(r)$ outside the star, and hence
from~(\ref{rhoD}) we see that,
\begin{equation}
     \tilde \rho = \frac{8\pi G r_H^2}{c^4}\rho_D=\frac{\Delta}{x^2}\quad (r>R_*).
\label{rescaled rho}
\end{equation}
Furthermore, since we are interested in the metric at sub-Hubble distances, $x=r/r_H\ll 1$, and in the weak field regime,
$x\gg r_S/r_H=3\Delta^2$, one can approximate $A(x)$ in all cases of interest  $A(x)\approx 1$ in
 the TOV equation~(\ref{TOV equation:2}).
In what follows we consider separately Cases I and II.

\subsection*{Case I: Field-like dark mass: $P_r=P, P_\perp=0$ }

In Case I and outside the star Eq.~(\ref{TOV equation:2}) can be simplified as,
\begin{equation}
  \frac{dp}{dx}+\bigg[\frac{2}x -x+\frac{\Delta}{x} +\frac{3}{2}\frac{\Delta^2}{x^2}\bigg]p+\frac{x}{2}p^2
                = \Big(x-\frac{\Delta}{2x}-\frac{3}{2}\frac{\Delta^2}{x^2}\Big)\frac{\Delta}{x^2}
 \,,
\label{TOV equation:3}
\end{equation}
where $\Delta=\sqrt{r_S/(3r_H)}$. Note that the terms $dp/dx+2p/x$ dominate the equation for any $x<1$
and therefore they determine the form of the solution,
\begin{equation}
 p = \frac{p_0}{x^2}
\,,
\label{form of p(x) Case I}
\end{equation}
where $p_0$ is an integration constant we wish to determine. Since~(\ref{form of p(x) Case I})
solves $dp/dx+2p/x=0$, the remaining terms in~(\ref{TOV equation:2}) combine to an algebraic equation
that also must be satisfied. We shall now show that this algebraic equation
determines $p_0$,
\begin{equation}
  \bigg(\!\!-\!x +\frac{\Delta}{x}+\frac{3}{2}\frac{\Delta^2}{x^2}\bigg)p_0+\frac{p_0^2}{2x}
                = \Big(x-\frac{\Delta}{2x}-\frac{3}{2}\frac{\Delta^2}{x^2}\Big)\Delta
 \,.
\label{TOV equation:4}
\end{equation}
At first sight this does not appear as a consistent equation since it contains a baroque $x$ dependence.
A closer look at~(\ref{TOV equation:4}) reveals however that it is consistent. To see that let us
split~(\ref{TOV equation:4}) into two equations and require that each of them be separately satisfied,
\begin{equation}
  -\bigg(\!x -\frac{\Delta}{2x}-\frac{3}{2}\frac{\Delta^2}{x^2}\bigg)p_0 = \Big(x-\frac{\Delta}{2x}-\frac{3}{2}\frac{\Delta^2}{x^2}\Big)\Delta
\,,\qquad \frac{\Delta}{2x}p_0+\frac{p_0^2}{2x} = 0
 \,.
\label{TOV equation:5}
\end{equation}
The first equation is solved for $p_0=-\Delta$ while the second equation is satisfied when $p_0=0$ or $p_0=-\Delta$,
implying that both equations are solved when,
\begin{equation}
p_0=-\Delta \;\Rightarrow \; P = -\rho
\,.
\label{Case I: common solution pressure}
\end{equation}
We emphasize that~(\ref{Case I: common solution pressure}) solves~(\ref{TOV equation:3}) for all $x$
satisfying, $\Delta^2\ll x \ll 1$
($r_S\ll r \ll r_H$), which is also the range in which the static metric~(\ref{global monopole metric:2}) is valid.
Note that the solution~(\ref{Case I: common solution pressure}) is imposed in different ranges of $x$ by different terms
in Eq.~(\ref{TOV equation:3}) or~(\ref{TOV equation:4}).
Indeed,  when  $1\gg x\gg \Delta^{1/2}$ ($r\gg (r_S r_H^3)^{1/4}$) the first
term in~(\ref{TOV equation:4}) enforces the solution~(\ref{Case I: common solution pressure}),
when $\Delta^{1/2}\gg x\gg \Delta$ ($(r_S r_H^3)^{1/4}\gg r\gg (r_S r_H)^{1/2}$) the third
term in~(\ref{TOV equation:4}) enforces~(\ref{Case I: common solution pressure}),
and finally when $\Delta^2\ll x\ll \Delta$ ($r_S\ll r\ll (r_S r_H)^{1/2}$) the second
term in~(\ref{TOV equation:4}) enforces~(\ref{Case I: common solution pressure}).
It is remarkable that {\it all of these terms} impose the {\it same} solution for $p_0$.
That is of course no coincidence and the structure of the dominant solution can be traced back to Eq.~(\ref{TOV equation}), 
from which we see that when $P_r=-\rho$, $P_r$ solves the simple equation, $P_r^\prime=-2P_r/r$ which is solved by $P_r=P_0/r^2$.  

 When~(\ref{Case I: common solution pressure})
is inserted into~(\ref{Grr:2}) one sees that, in the exterior of a star, the negative pressure contribution
cancels the  $\Delta$ dependent contribution from $2GM(r)/c^2=\Delta\times r +2GM_*/c^2$ and one obtains,
\begin{equation}
 {\rm e}^{2\alpha(r)} = 1 - \Delta  -\frac{\Lambda}{3}r^2 - \frac{2GM_*}{c^2r} = {\rm e}^{-2\beta(r)}
\,,\qquad (r>R_*)
\,.
\label{Case I: alpha}
\end{equation}
This solution is valid everywhere in the exterior of a star in the weak field regime, $r_S\ll r \ll r_H$
and it is equivalent to the metric of a global monopole in the exterior of the monopole core.
Next we consider Case II, in which dark mass is assumed to be particle-like.

\subsection*{Case II: Particle-like dark mass: $P_r=P = P_\perp$ }

 In this fully isotropic case the TOV equation~(\ref{TOV equation:2}) reduces to
a Riccati equation ($|A(x)-1|\ll 1$),
\begin{equation}
  \frac{dp}{dx}+\bigg[\!-\!x +\frac{\Delta}{x}+\frac{3}{2}\frac{\Delta^2}{x^2}\bigg]p(x)
       +\frac{x}{2}p^2 = \Big(x-\frac{\Delta}{2x}-\frac{3}{2}\frac{\Delta^2}{x^2}\Big)\frac{\Delta}{x^2}
 \,.
\label{Case II: TOV equation}
\end{equation}
The simplest way to solve a Riccati equation for $p=p(x)$,
\begin{equation}
 p' + q_1(x) p + q_2(x) p^2 = q_0(x)
\label{Riccati original}
\end{equation}
is to introduce substitutions,
\begin{equation}
 Q_1=q_1-\frac{q_2^\prime}{q_2} = \!-\!x -\frac{1\!-\!\Delta}{x}+\frac{3}{2}\frac{\Delta^2}{x^2}
\,,\quad
 Q_0=-q_0q_2 = -\frac{\Delta}{2}+\frac{\Delta^2}{4x^2}+\frac{3}{4}\frac{\Delta^3}{x^3}
\,,\quad
 p = \frac{u^\prime}{q_2 u}
\,,\quad
\nonumber
\end{equation}
upon which Eq.~(\ref{Riccati original}) reduces to a second order linear differential equation,
\begin{equation}
 u^{\prime\prime} + Q_1 u^\prime + Q_0u = 0
\,.
\label{Riccati original:2}
\end{equation}
A further substitution, $v = a(x) u$, with  $a^\prime/a= Q_1(x)/2$, reduces~(\ref{Riccati original:2}) to,
\begin{equation}
 v^{\prime\prime} + \bigg[\!-\!\frac{a^{\prime\prime}}{a}+Q_0\bigg ] v = 0
\,,
\label{Riccati original:3}
\end{equation}
which in our case becomes,
\begin{equation}
 v^{\prime\prime} +  \bigg[-\frac{x^2}{4}+\frac{3\Delta^2}{4x}-\frac{3\!-\!4\Delta}{4x^2}
                      +\frac{9\Delta^2}{4x^3}-\frac{9\Delta^4}{16x^4}\bigg] v= 0
\,.
\label{Riccati our Case II}
\end{equation}
Notice that once we know $v=v(x)$, then the pressure can be calculated by,
\begin{equation}
 p = \frac{2}{x}\left[-\frac{Q_1}{2}+\frac{v^\prime}{v}\right]
    = 1 +\frac{1-\Delta}{x^2}-\frac{3}{2}\frac{\Delta^2}{x^3}+\frac{2}{x}\frac{v^\prime}{v}
\,.
\label{Case II: pressure}
\end{equation}

Let us solve~(\ref{Riccati our Case II}) for $v(x)$. A careful look at all terms in~(\ref{Riccati our Case II})
reveals that it is the fourth term ($\propto 1/x^2$) inside the square brackets
that dominates for the relevant range of coordinates, $\Delta^2\ll x\ll 1$ ($r_S\ll r \ll r_H$)
However, that does not mean that one can neglect other terms.
It is hard to solve~(\ref{Riccati our Case II}) in full generality. Nevertheless, one can find an approximate solution as follows.
Let us first rewrite~(\ref{Riccati our Case II}) as,
\begin{eqnarray}
v^{\prime\prime} + \bigg(\frac{F_2}{x^2}+\frac{F_3}{x^3}-\frac{F_4}{x^4}\bigg) v \!&=&\! 0
\,,\quad
F_2(x) =-\frac{x^4}{4}+\frac{3\Delta^2}{4}x-\frac{3\!-\!4\Delta}{4}
\nonumber\\
&&\hskip 1.4cm
 F_3  = \frac{9\Delta^2}{4}
\,,\qquad
 F_4=
\bigg(\frac{3\Delta^2}{4}\bigg)^2
\,,\qquad
\label{Riccati our Case II:2}
\end{eqnarray}
where $F_2(x)$ is an adiabatic function of $x$ on the whole interval, $\Delta^2\ll x\ll 1$,
in the sense that, $F_2^\prime/F_2\ll F_2$
(attempting to include $F_3/x$ and/or $F_4/x^2$ into $F_2$ would break adiabaticity when $\Delta^2\ll x\ll \Delta$).
The following substitutions,
$v(x)\rightarrow v(y(x))$ with $y=1/x$ and $w(y) = y v(y)$ reduce~(\ref{Riccati our Case II:2})
to the Whittaker differential equation,
\begin{equation}
 \frac{d^2w}{dy^2} + \bigg(\frac{F_2}{y^2}+\frac{F_3}{y}-F_4\bigg)w(y) = 0
\,,
\label{Whittaker equation}
\end{equation}
whose two linearly independent solutions are given by the Whittaker functions,
\begin{equation}
 w(y) \sim M_{\frac{F_3}{2\sqrt{F_4}}\,,\,\frac12\sqrt{1-4F_2}}\Big(2\sqrt{F_4}y\Big)\,,
\quad W_{\frac{F_3}{2\sqrt{F_4}}\,,\,\frac12\sqrt{1-4F_2}}\Big(2\sqrt{F_4}y\Big)
\,.
\label{Whittaker functions}
\end{equation}
These functions are related to the confluent hypergeometric function by standard relations,
\begin{eqnarray}
 M_{\nu,\mu}(z)  \!&=&\! {\rm e}^{-z/2}z^{\frac12+\mu}\times_{1}F_1\bigg(\frac12+\mu-\nu;1+2\mu;z\bigg)
\nonumber\\
 W_{\nu,\mu}(z)  \!&=&\! {\rm e}^{-z/2}z^{\frac12-\mu}\frac{\Gamma(2\mu)}{\Gamma(\frac12+\mu-\nu)}
                  \times_{1}F_1\bigg(\frac12-\mu-\nu;1-2\mu;z\bigg)
\nonumber\\
\!&+&\! {\rm e}^{-z/2}z^{\frac12+\mu}\frac{\Gamma(-2\mu)}{\Gamma(\frac12-\mu-\nu)}
                   \times_{1}F_1\bigg(\frac12+\mu-\nu;1+2\mu;z\bigg)
\,.
\label{Whittaer to confluent}
\end{eqnarray}
By making use of~(\ref{Whittaer to confluent}) and~(\ref{Whittaker functions})
one gets two lineraly independent solutions for $v(x) = xw(1/x)$,
\begin{eqnarray}
 v_\pm(x) \!&\sim&\! {\rm e}^{-3\Delta^2/(4x)}x^{\frac12(1\mp\delta_v)}
        \times_{1}\!F_1\bigg(\!\!-1\!\pm\frac{\delta_v}{2};1\pm\delta_v;\frac{3\Delta^2}{2x}\bigg)
\,,
\label{confluent hypergeometric function:+-}\\
\label{confluent hypergeometric function:-}
\end{eqnarray}
where we have defined,
\begin{equation}
 \delta_v\equiv \sqrt{1-4F_2}=\sqrt{4(1\!-\!\Delta)+x^4-3x\Delta^2}\approx 2\sqrt{1-\Delta}+\frac{x^4}{8\sqrt{1-\Delta}}
\,.
\label{delta v def}
\end{equation}
The physical solution is a linear combination of $v_+$ and $v_-$ in~(\ref{confluent hypergeometric function:+-}),
\begin{equation}
v=C\big[v_+(x)+cv_-(x)\big],
\label{v solution}
\end{equation}
where $C,c$ are unknown (integration) constants.
When~(\ref{v solution}) is inserted into~(\ref{Case II: pressure}) one obtains,
\begin{equation}
 p = 1 + \frac{1-\Delta}{x^2}-\frac{3\Delta^2}{2x^3} 
           + \frac{1}{1+cx^{\delta_v}}
                           \bigg\{\bigg[\frac{1-\delta_v}{x^2}+\frac{3\Delta^2}{2x^3}\frac{3}{1+\delta_v}\bigg]
                              + cx^{\delta_v}\bigg[\frac{1+\delta_v}{x^2}+\frac{3\Delta^2}{2x^3}\frac{3}{1-\delta_v}\bigg]
                         \bigg\}
\,,
\label{Case II: pressure:2}
\end{equation}
where we have neglected derivatives of $\delta_v(x)$ (which is justified in the adiabatic approximation we are using).
One way of determining which is the case is to solve for pressure inside the star and then continuously match at the exterior solution.
A detailed analysis~\footnote{The TOV equation can be solved inside the star ($x<R_*/r_H$). One can show
that the approximate solution for pressure is of the form,
\[
p(x) = -\frac{5}{9}\tilde \rho_* x + p_0 +{\rm higher\; order\; terms}\,,
\]
where $p_0\ll \tilde \rho_*$ is the pressure at the center of the star (since the star is assumed to be non-relativistic, it is
reasonable to assume that the pressure at the center of the star is much less than the energy density) and
$\tilde \rho_* = 3r_H^2r_S/R_*^3=8\phi_*^3/(3\Delta^4)\gg 1$ is the rescaled energy density~(\ref{rhoB})
and we introduced the dimensionless potential at the star surface, $\phi_*\equiv r_S/(2R_*)$. For our Sun $\phi_*\sim 10^{-6}$
and $\Delta\sim 3\times 10^{-11}\ll \phi_*$. To get the pressure at the star surface, we need to compare
$\Delta/x_*=2\phi_*/(3\Delta)\gg 1$ with $-x_*p(x_*)\simeq 10\phi_*/3 -4(p_0/\rho_*)\phi_*^2\ll 1$
and hence the pressure contribution just outside the star must be negligible when compared to $\Delta/x^2$.
This means that the $v_+$ contribution in~(\ref{v solution}--\ref{Case II: pressure:2}) must dominate, thus justifying
Eq.~(\ref{p = p+}).
}
shows that $v_+$ dominates, {\it i.e.} $|c|x^{\delta_v}\ll 1$ and Eq.~(\ref{Case II: pressure:2}) evaluates to, 
\begin{equation}
p = 1 + \frac{x^2}{8}+\frac{\Delta^3}{2x^3}  
\,.
\label{p = p+}
\end{equation}
With this in mind and upon inserting~(\ref{p = p+}) into~(\ref{Grr:2}) one obtains the following (approximate)
equation for the metric outside a star ($r>R_*$),
\begin{equation}
 \frac{d\alpha}{dx} =\frac{1}{2}\frac{d}{dx}\ln\bigg(1-\Delta-x^2-\frac{3\Delta^2}{x}\bigg)
               +\frac12\bigg(\frac{\Delta}{x}+x+\frac{x^3}{8}+\frac{\Delta^3}{2x^2}\bigg)
\,.
\label{Case II: alpha}
\end{equation}
Let us focus on the terms in the second parentheses. The second term changes (halfs) the cosmological constant while the third term
is a small corrections that is negligible except at very large distances and the last (third) term is 
a small (order $\Delta$) correction to the Newtonian term. Upon neglecting the last two terms, 
Eq.~(\ref{Case II: alpha}) can be integrated to obtain,
\begin{equation}
{\rm e}^{2\alpha(r)} = \left(1-\Delta -\frac{\Lambda}{6}r^2-\frac{2GM}{r}\right)\left(\frac{r}{r_H}\right)^{\Delta}
\,.
\label{Case II: alpha:2}
\end{equation}
This result is used in the main text to write the metric tensor~(\ref{global monopole metric}).

\bibliography{mybibfile}

\section*{References}
\begin{thebibliography}{99}


\bibitem{Verlinde:2016toy}
  E.~P.~Verlinde,
  ``Emergent Gravity and the Dark Universe,''
  arXiv:1611.02269 [hep-th].

\bibitem{Marunovic:2014hla}
  A.~Marunovic and T.~Prokopec,
  ``Global monopoles can change Universe's topology,''
  Phys.\ Lett.\ B {\bf 756} (2016) 268
  doi:10.1016/j.physletb.2016.03.030
  [arXiv:1411.7402 [gr-qc]].

\bibitem{Barriola:1989hx}
  M.~Barriola and A.~Vilenkin,
  ``Gravitational Field of a Global Monopole,''
  Phys.\ Rev.\ Lett.\  {\bf 63} (1989) 341.
  doi:10.1103/PhysRevLett.63.341

\bibitem{Marunovic:2015wha}
  A.~Marunovic and T.~Prokopec,
  ``Topological inflation with graceful exit,''
  JCAP {\bf 1604} (2016) no.04,  052
  doi:10.1088/1475-7516/2016/04/052
  [arXiv:1508.05010 [gr-qc]].

\bibitem{Marunovic:2013eka}
  A.~Marunović and M.~Murković,
  ``A novel black hole mimicker: a boson star and a global monopole nonminimally coupled to gravity,''
  Class.\ Quant.\ Grav.\  {\bf 31} (2014) 045010
  doi:10.1088/0264-9381/31/4/045010
  [arXiv:1308.6489 [gr-qc]].

\bibitem{Spagna:2016jnz}
  A.~Spagna, M.~Crosta, M.~G.~Lattanzi and P.~Re Fiorentin,
  ``The Gaia mission: the dawn of Astrometric Cosmology? Status and prospects after 14 months of science operations,''
  J.\ Phys.\ Conf.\ Ser.\  {\bf 718} (2016) no.3,  032005.
  doi:10.1088/1742-6596/718/3/032005

\bibitem{GAIA_ESA}
http://www.cosmos.esa.int/web/gaia.

\bibitem{Hees:2015ixa}
  A.~Hees, D.~Hestroffer, C.~Le Poncin-Lafitte and P.~David,
  ``Tests of gravitation with Gaia observations of Solar System Objects,''
  arXiv:1509.06868 [gr-qc].



\bibitem{Doeleman:2009te}
  S.~Doeleman {\it et al.},
  ``Imaging an Event Horizon: submm-VLBI of a Super Massive Black Hole,''
  arXiv:0906.3899 [astro-ph.CO].

\bibitem{EHT}
 http://www.eventhorizontelescope.org/.

\bibitem{Wambsganss:2016qpg}
  J.~Wambsganss,
  ``Discovering Extrasolar Planets with Microlensing Surveys,''
  ASP Conf.\ Ser.\  {\bf 505} (2016) 35.

\bibitem{Rahvar:2015eba}
  S.~Rahvar,
  ``Gravitational microlensing I: A unique astrophysical tool,''
  Int.\ J.\ Mod.\ Phys.\ D {\bf 24} (2015) no.07,  1530020.
  doi:10.1142/S0218271815300207

\bibitem{Mao:2012za}
  S.~Mao,
  ``Astrophysical Applications of Gravitational Microlensing,''
  Res.\ Astron.\ Astrophys.\  {\bf 12} (2012) 947
  doi:10.1088/1674-4527/12/8/005
  [arXiv:1207.3720 [astro-ph.GA]].

\bibitem{Brouwer:2016dvq}
  M.~M.~Brouwer {\it et al.},
  ``First test of Verlinde's theory of Emergent Gravity using Weak Gravitational Lensing measurements,''
  doi:10.1093/mnras/stw3192
  arXiv:1612.03034 [astro-ph.CO].

\bibitem{Iorio:2016qzr}
  L.~Iorio,
  ``Are we close to put the anomalous perihelion precessions from Verlinde's emergent gravity to the test?,''
  arXiv:1612.03783 [gr-qc].

\end{thebibliography}

\end{document}